\documentclass[journal]{IEEEtran}

\usepackage{amsmath,amsfonts,bm}

\def\1{\bm{1}}

\def\p{{\mathbf{p}}}

\def\D{{\mathbf{D}}}

\def\I{{\mathbf{I}}}

\def\L{{\mathbf{L}}}

\def\P{{\mathbf{P}}}

\def\U{{\mathbf{U}}}

\def\W{{\mathbf{W}}}

\DeclareMathAlphabet{\mathsfit}{\encodingdefault}{\sfdefault}{m}{sl}
\SetMathAlphabet{\mathsfit}{bold}{\encodingdefault}{\sfdefault}{bx}{n}

\def\gF{{\mathcal{F}}}

\usepackage{amsmath,amsopn,amssymb}
\usepackage{graphicx,xspace,color,soul}
\usepackage{epsfig,subfigure}
\usepackage{longtable,multirow}
\usepackage{tabularx}
\usepackage{array,float}
\usepackage{cite}
\usepackage{algorithmic}
\usepackage[linesnumbered, ruled]{algorithm2e}
\usepackage{bm,epstopdf}
\usepackage{caption2}
\newcolumntype{Y}{>{\centering\arraybackslash}X}

\def\ie{{\textit{i.e.}}}
\def\eg{{\textit{e.g.}}}
\def\etc{{\textit{etc}}}
\def\etal{{\textit{et al.~}}}

\def\p{{\mathbf p}}

\def\d{{\mathbf d}}
\def\f{{\mathbf f}}

\def\n{{\mathbf n}}

\def\v{{\mathbf v}}
\def\w{{\mathbf w}}

\def\D{{\mathbf D}}
\def\I{{\mathbf I}}
\def\L{{\mathbf L}}

\def\I{{\mathbf I}}
\def\U{{\mathbf U}}
\def\W{{\mathbf W}}

\def\cM{{\mathcal M}}

\begin{document}

\title{Dynamic Point Cloud Denoising via Manifold-to-Manifold Distance}

\author{
    Wei~Hu,~\IEEEmembership{Member,~IEEE,}
	Qianjiang~Hu,~\IEEEmembership{Student Member,~IEEE,}
	Zehua~Wang,~\IEEEmembership{Student Member,~IEEE,}
	and~Xiang~Gao,~\IEEEmembership{Student Member,~IEEE}
	\thanks{W. Hu, Q. Hu, Z. Wang and X. Gao are with Wangxuan Institute of Computer Technology, Peking University, No. 128, Zhongguancun North Street, Beijing, China. E-mail: \{forhuwei, hqjpku, klsl1307, gyshgx868\}@pku.edu.cn. Corresponding author: Wei Hu. }

}


\maketitle

\begin{abstract}
3D dynamic point clouds provide a natural discrete representation of real-world objects or scenes in motion, with a wide range of applications in immersive telepresence, autonomous driving, surveillance, \etc. 
Nevertheless, dynamic point clouds are often perturbed by noise due to hardware, software or other causes. 
While a plethora of methods have been proposed for static point cloud denoising, few efforts are made for the denoising of dynamic point clouds, which is quite challenging due to the irregular sampling patterns both spatially and temporally. In this paper, we represent dynamic point clouds naturally on spatial-temporal graphs, and exploit the temporal consistency with respect to the underlying surface (manifold). 
In particular, we define a manifold-to-manifold distance and its discrete counterpart on graphs to measure the variation-based intrinsic distance between surface patches in the temporal domain, provided that graph operators are discrete counterparts of functionals on Riemannian manifolds. Then, we construct the spatial-temporal graph connectivity between corresponding surface patches based on the temporal distance and between points in adjacent patches in the spatial domain. 
Leveraging the initial graph representation, we formulate dynamic point cloud denoising as the joint optimization of the desired point cloud and underlying graph representation, regularized by both spatial smoothness and temporal consistency. 
We reformulate the optimization and present an efficient algorithm. 
Experimental results show that the proposed method significantly outperforms independent denoising of each frame from state-of-the-art static point cloud denoising approaches, on both Gaussian noise and simulated LiDAR noise.          

\end{abstract}


\vspace{-0.05in}
\section{Introduction}
\label{sec:intro}

\begin{figure}[htbp]
     \centering
     \includegraphics[width=0.75\linewidth]{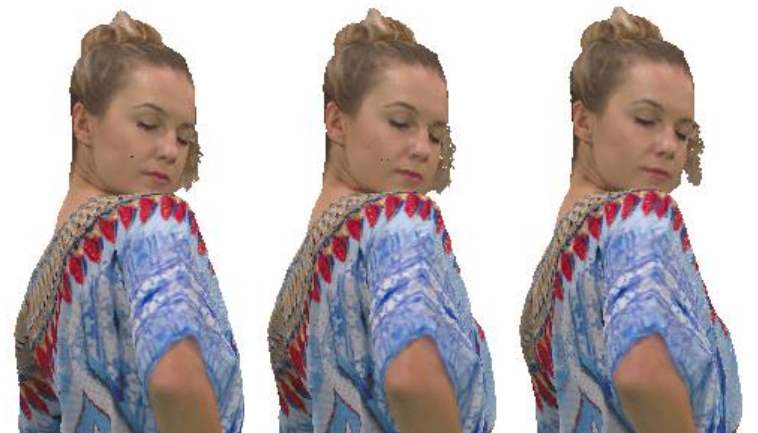}
     \caption{Three frames ($1203$, $1204$, $1205$) in the dynamic point cloud sequence \textit{Longdress} \cite{MPEG}. We observe the temporal consistency in geometry along the time dimension. }
     \label{fig:noisy}
 \end{figure}

The maturity of depth sensing, laser scanning\footnote{Commercial products include Microsoft Kinect (2010-2014), Intel RealSense (2015-), Velodyne LiDAR (2007-2020), LiDAR scanner of Apple iPad Pro (2020), {\it etc.}} and image processing has enabled convenient acquisition of 3D dynamic point clouds, a natural representation for arbitrarily-shaped objects varying over time \cite{Rusu20113DIH}. 
A dynamic point cloud consists of a sequence of static point clouds, each of which is composed of a set of points irregularly sampled from the continuous surfaces of objects or scenes. Each point has geometry information (\ie, 3D coordinates) and possibly attribute information such as color, as shown in Fig.~\ref{fig:noisy}. 
Dynamic point clouds have been widely applied in various applications, such as augmented and virtual reality \cite{burdea2003virtual,schmalstieg2016augmented}, autonomous driving \cite{ChenLFVW:20}, surveillance and monitoring \cite{benedek20143d}. 

Point clouds are often contaminated by noise, which comes from hardware, software or other causes. Hardware wise, noise occurs due to the inherent limitations of the acquisition equipment. Software wise, when reconstructing point clouds from images, points may locate somewhere completely wrong due to matching ambiguities or imprecise triangulation. 
The noise perturbation often significantly affects the analysis of point clouds since the underlying structures are deformed. 
Hence, dynamic point cloud denoising is crucial to relevant applications. 
However, it is quite challenging to address, due to the irregular sampling patterns in each frame and possibly varying number of points in different frames, which also means there is no explicit temporal correspondence between points. 

While a plethora of approaches have been proposed for static point cloud denoising \cite{Alexa2003Computing,Guennebaud2007Algebraic,A2009Feature,Huang2013Edge,Hui2009Consolidation,Lipman2007Parameterization,avron,Mattei2017Point,Dinesh2018Fast,dinesh18arxiv,zeng20183d,duan2018weighted,Hu19tsp,NeuralProj2019, PCN2020, TotalDenoising2019, luo2020differentiable}, few efforts are made for the denoising of \textit{dynamic} point clouds in the literature. 
Whereas it is possible to apply existing static point cloud denoising methods to each frame of a dynamic point cloud sequence independently, the \textit{inter-frame correlation} would be neglected, which may also lead to inconsistent denoising results in the temporal domain. 

To this end, we propose to leverage the spatial-temporal correlation for dynamic point cloud denoising, particularly exploiting the temporal consistency between corresponding local surface patches based on the proposed manifold-to-manifold distance.   
Since dynamic point clouds are irregular both spatially and temporally, we represent them naturally on {\it spatial-temporal graphs}, where
a node represents a sampled point and the graph connectivity characterizes the spatial-temporal correlation in the point cloud. 
In particular, as point clouds are discrete samples of functions on Riemannian manifolds and often exhibit piece-wise smoothness \cite{Hu2020gsp}, 
we assume that the low-dimensional embedding of dynamic point clouds lies on {\it smooth manifolds} both spatially and temporally, and exploit the spatial smoothness and the temporal consistency based on the graph representation.  

The key is to exploit the temporal consistency between surface patches that correspond to the same local surface of the observed 3D object or scene, which significantly differs from spatial smoothness among adjacent points exploited in previous works \cite{zeng20183d,Hu19tsp}, as there is no explicit point-to-point correspondence between point cloud frames over time. 
Even if the underlying 2D manifold is static in two frames, different irregular sampling patterns would lead to different samples in the two frames, resulting in nonzero conventional measurements such as the Euclidean distance. 
Hence, it is crucial to define a temporal distance to intrinsically measure the distance between discrete surface patches in adjacent point cloud frames with respect to the underlying continuous manifold.  

As point clouds are discrete samples from continuous manifolds, we first propose a {\it manifold-to-manifold distance} that measures the difference in the variation of manifolds, which takes the second-order differential of normal coordinates via the Laplace-Beltrami operator. 
Then, we derive the {\it discrete counterpart} of the manifold-to-manifold distance, based on that the random walk graph Laplacian is a discrete approximation to the weighted Laplace-Beltrami operator under certain constraints \cite{ting2010analysis,hein2007graph}.
This leads to the proposed temporal distance, which measures the difference in the variation of surface patches by operating the random walk graph Laplacian on the normal coordinates of points. Based on the derived distance, we propose a temporal matching method to infer temporally corresponding surface patches in adjacent point cloud frames. 
The temporal graph topology is then constructed to connect each pair of {\it temporally corresponding patches}, whereas the spatial graph connectivity captures the similarity between each pair of spatially adjacent points in the same frame. 

Based on the spatial-temporal graph connectivity initially constructed from the noisy observation, we formulate dynamic point cloud denoising as the joint optimization of the desired point cloud and underlying spatial-temporal graph representation. 
To exploit the spatial smoothness and temporal consistency, we regularize the formulation with respect to the underlying spatial-temporal graph by  1) smoothness of adjacent patches in the current frame via the graph Laplacian regularizer \cite{Shuman13} and 2) a defined local temporal consistency term between corresponding patches.   
Finally, we propose an efficient algorithm to solve the problem formulation.
We design an alternating minimization algorithm to optimize the underlying frame and spatial-temporal graph alternately. 
When the graph is initialized or fixed, we update the underlying frame via a closed-form solution.  
When the underlying frame is updated, the optimization of the temporal graph is reformulated as a linear program, while that of the spatial graph is cast into a feature graph learning problem as in \cite{Hu19tsp}. 
This process is iterated until the convergence of the objective value. 
Experimental results show that the proposed method outperforms independent denoising of each frame from state-of-the-art static point cloud denoising approaches over nine widely employed dynamic point cloud sequences, on both Gaussian noise and simulated LiDAR noise. 

To summarize, the main contributions of our work include:
\begin{itemize}
    \item We propose dynamic point cloud denoising by exploiting the temporal consistency with respect to the underlying manifold, based on the proposed manifold-to-manifold distance and its discrete counterpart on graphs to measure the intrinsic distance between local surface patches in the temporal domain.  
    
    \item We formulate dynamic point cloud denoising as the joint optimization of the desired point cloud and underlying spatial-temporal graph representation, regularized by the spatial smoothness as well as the defined local temporal consistency between temporally corresponding patches.
    
    \item We present an efficient algorithm to optimize the point cloud and spatial-temporal graph representation alternately, where we reformulate the temporal graph learning and spatial graph learning respectively, leading to superior denoising performance.  
\end{itemize}

The outline of this paper is as follows. We first review previous static point cloud denoising methods in Section~\ref{sec:related}. 
Then we elaborate on the proposed manifold-to-manifold distance and its discrete counterpart for temporal matching in dynamic point clouds in Section~\ref{sec:temporal}. 
Next, we present the proposed problem formulation in Section~\ref{sec:method} and discuss the algorithm development in Section~\ref{sec:algorithm}.
Finally, experimental results and conclusions are presented in Section~\ref{sec:results} and Section~\ref{sec:conclude}, respectively.

\vspace{-0.05in}
\section{Related Work}
\label{sec:related}

To the best of our knowledge, there are few efforts on dynamic point cloud denoising in the literature. 
\cite{Schoenenberger2015Graph} provides a short discussion on how the proposed graph-based static point cloud denoising naturally generalizes to time-varying inputs such as 3D dynamic point clouds.  
Therefore, we discuss previous works on \textit{static} point cloud denoising, which can be divided into six classes: moving least squares (MLS)-based methods,
locally optimal projection (LOP)-based methods, sparsity-based methods, non-local methods, graph-based methods, and deep-learning based methods. 

\textbf{MLS-based methods.} Moving Least Squares (MLS)-based methods aim to approximate a smooth surface from the input point cloud and minimize the geometric error of the approximation. 
Alexa {\it et al.} approximate with a polynomial function on a local reference domain to best fit neighboring points in terms of MLS \cite{Alexa2003Computing}. 
Other similar solutions include algebraic point set surfaces (APSS) \cite{Guennebaud2007Algebraic} and robust implicit MLS (RIMLS) \cite{A2009Feature}. 
However, this class of methods may lead to over-smoothing results. 

\textbf{LOP-based methods.} Locally Optimal Projection (LOP)-based methods also employ surface approximation for denoising point clouds. 
Unlike MLS-based methods, the operator is non-parametric, which performs well in cases of ambiguous orientation. 
For instance, Lipman {\it et al.} define a set of points that represent the estimated surface by minimizing the sum of Euclidean distances to the data points \cite{Lipman2007Parameterization}. 
The two branches of \cite{Lipman2007Parameterization} are weighted LOP (WLOP) \cite{Hui2009Consolidation} and anisotropic WLOP (AWLOP) \cite{Huang2013Edge}. \cite{Hui2009Consolidation} produces a set of denoised, outlier-free and more evenly distributed particles over the original dense point cloud to keep the sample distance of neighboring points. 
\cite{Huang2013Edge} modifies WLOP with an anisotropic weighting function so as to preserve sharp features better. Nevertheless, LOP-based methods may also over-smooth point clouds.

\textbf{Sparsity-based methods.} These methods are based on sparse representation of point normals.
Regularized by sparsity, a global minimization problem is solved to obtain sparse reconstruction of point normals.
Then the positions of points are updated by solving another global $l_0$ \cite{Sun2015Denoising} or $l_1$ \cite{avron} minimization problem based on a local planar assumption.
Mattei {\it et al.} \cite{Mattei2017Point} propose \textit{Moving Robust Principal Components Analysis} (MRPCA) approach to denoise 3D point clouds via weighted $l_1$ minimization to preserve sharp features.
However, when locally high noise-to-signal ratios yield redundant features, these methods may not perform well and lead to over-smoothing or over-sharpening \cite{Sun2015Denoising}.

\textbf{Non-local methods.} Inspired by non-local means (NLM) \cite{buades2005non} and BM3D \cite{4271520} image denoising algorithms, this class of methods exploit self-similarities among surface patches in a point cloud. 
Digne {\it et al.} utilize a NLM algorithm to denoise static point clouds \cite{digne2012similarity}, while Rosman {\it et al.} deploy a BM3D method \cite{rosman2013patch}. 
Deschaud {\it et al.} extend the non-local denoising (NLD) algorithm for point clouds, where the neighborhood of each point is described by the polynomial coefficients of the local MLS surface to compute point similarity \cite{deschaud10}.
\cite{sarkar2018structured} utilizes patch self-similarity and optimizes for a low rank (LR) dictionary representation of the extracted patches to smooth 3D patches.
Nevertheless, the computational complexity of these methods is usually high.

\textbf{Graph-based methods.}~~~This family of methods represent a point cloud on a graph, and design graph filters for denoising.
Schoenenberger {\it et al.} \cite{Schoenenberger2015Graph} construct a $K$-nearest-neighbor graph on the input point cloud and then formulate a convex optimization problem regularized by the gradient of the point cloud on the graph.
Dinesh {\it et al.} \cite{dinesh18arxiv} design a reweighted graph Laplacian regularizer for surface normals, which is deployed to formulate an optimization problem with a general $\ell p$-norm fidelity term that can explicitly model two types of independent noise.  
Zeng {\it et al.} \cite{zeng20183d} propose a low-dimensional manifold model (LDMM) with graph Laplacian regularization (GLR) and exploit self-similar surface patches for denoising. 
Instead of directly smoothing the 3D coordinates or surface normals, Duan {\it et al.} \cite{duan2018weighted} estimate the local tangent plane of each point based on a graph, and then reconstruct 3D point coordinates by averaging their projections on multiple tangent planes.
Hu {\it et al.} \cite{Hu19tsp} propose feature graph learning to optimize edge weights given a single or even partial observation assumed to be smooth with respect to the graph.  
However, the temporal dependency is not exploited yet in this class of methods. 

Our method extends the previous work \cite{Hu19tsp} to dynamic point cloud denoising, by exploiting the temporal consistency that is significantly different from spatial smoothness. 
While \cite{Hu19tsp} leverages smoothness among adjacent patches in the spatial domain, we further consider the temporal consistency between patches that correspond to the same underlying manifold over time but are sampled with different irregular patterns, which is the key challenge of dynamic point cloud denoising.  

\textbf{Deep-learning based methods.}
With the advent of neural networks for point clouds, deep-learning based point cloud denoising has received increasing attention. 
Among them, Neural Projection \cite{NeuralProj2019} leverages PointNet \cite{qi2017pointnet} to predict the tangent plane at each point, and projects points to the tangent planes. 
PointCleanNet \cite{PCN2020} predicts displacement of points from the clean surface via PCPNet \cite{guerrero2018pcpnet}---a variant of PointNet, which is trained end-to-end by minimizing the $\ell 2$ distance between the denoised point cloud and the ground truth. 
While the aforementioned methods require the supervision of the ground truth, Total Denoising \cite{TotalDenoising2019} is the first unsupervised deep learning method for point cloud denoising, assuming that points with denser surroundings are closer to the underlying surface. 
However, these methods are not designated to recover the underlying surface explicitly, which are often sensitive to outliers and may lead to point cloud shrinking. 
Luo \etal propose to learn the underlying surface (manifold) explicitly \cite{luo2020differentiable}, by sampling a subset of points with low noise via differentiable pooling and then reconstructing the underlying manifold from these points and their embedded neighborhood features.

\section{Manifold-to-Manifold Distance for Temporal Matching in Point Clouds}
\label{sec:temporal}
In this section, we first provide the preliminaries in basic graph concepts, and then propose a manifold-to-manifold distance given that point clouds serve as a discrete representation of underlying continuous manifolds over a set of sampled nodes. 
Based on the manifold-to-manifold distance, we derive its discrete counterpart and present a temporal matching method for dynamic point clouds. 

\subsection{Preliminaries}
We represent dynamic point clouds on undirected graphs. An undirected graph $\mathcal{G}=\{\mathcal{V},\mathcal{E},\mathbf{A}\}$ is composed of a node set $\mathcal{V}$ of cardinality $\left|\mathcal{V}\right|=N$, an edge set $\mathcal{E}$ connecting nodes, and a weighted adjacency matrix $\mathbf{A}$. $\mathbf{A} \in \mathbb{R}^{N \times N}$ is a real and symmetric matrix, where $a_{i,j} \geq 0$ is the weight assigned to the edge $(i,j)$ connecting nodes $i$ and $j$. Edge weights often measure the similarity between connected nodes. 

The graph Laplacian matrix is defined from the adjacency matrix. Among different variants of Laplacian matrices, the commonly used \textit{combinatorial graph Laplacian} \cite{chung1997spectral,hu15tip} is defined as $ \mathbf{L}:=\mathbf{D}-\mathbf{A} $, where $ \mathbf{D} $ is the \textit{degree matrix}---a diagonal matrix where $ d_{i,i} = \sum_{j=1}^N a_{i,j} $. 
A random walk Laplacian is defined as $ \mathbf{L}_{\text{rw}}:=\D^{-1}\L=\I-\D^{-1}\mathbf{A} $, which normalizes the degree of each node to $1$. 

Graph signal refers to data that resides on the nodes of a graph. In our scenario, the coordinates and normals of each point in the dynamic point cloud are the graph signal.

\begin{figure}[t]
  \begin{center}
    \begin{tabular}{c}
    \includegraphics[width=0.49\textwidth]{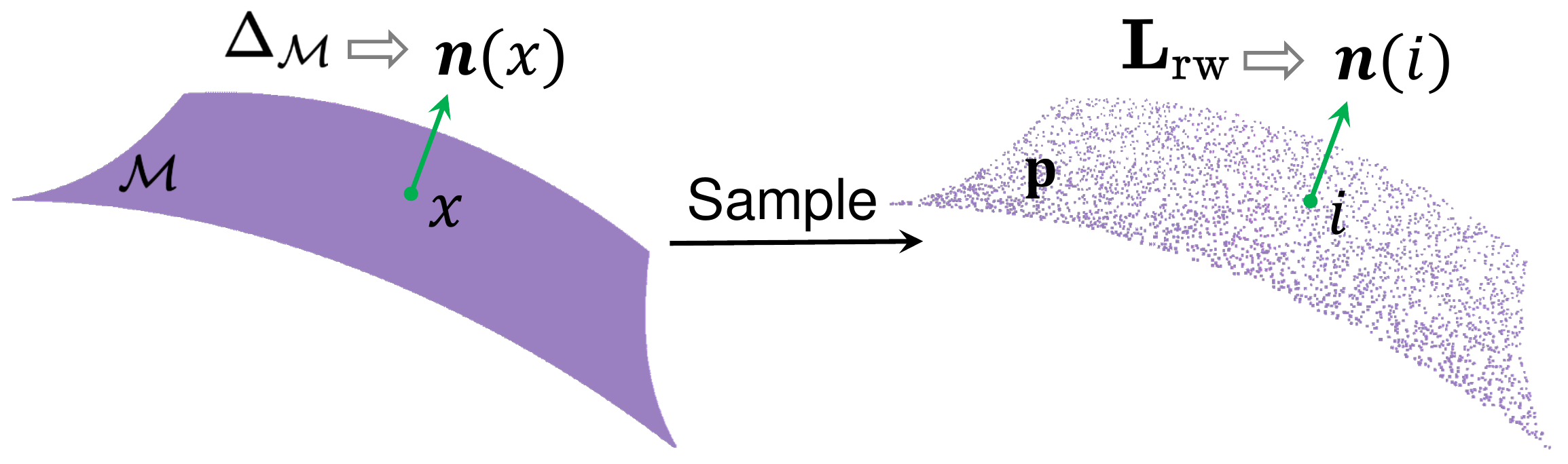}
  \end{tabular}
\end{center}
\vspace{-0.1in}
\caption{Demonstration of the distance measure for a manifold and its discrete counterpart for a sampled point cloud.}
\label{fig:manifold_point}
\end{figure}

\subsection{Manifold-to-Manifold Distance}
\label{subsec:m2m_distance}
In dynamic point cloud filtering, the key to exploit the temporal information is to define a distance between frames of point clouds. 
As discussed in Section~\ref{sec:intro}, dynamic point clouds are irregularly sampled both spatially and temporally, resulting in no explicit temporal point-to-point correspondence for distance calculation. 

In order to explore local characteristics of point clouds, we measure the temporal distance based on \textit{patches}, each of which consists of a center point and its $K$ nearest neighboring points. 
Each patch is discrete sampling of a Riemannian manifold, which describes the 2D surface of a 3D object or scene. 
The temporal distance between two patches is challenging to define, since irregular sampling in each frame may lead to different sampling points even for a static underlying manifold. 

Given local surface patches $\p_l$ and $\p_m$ in two point cloud frames respectively, a desirable temporal distance $d(\p_l,\p_m)$ between the two surface patches should satisfy  
\begin{itemize}
    \item \textbf{Property 1}: $d(\p_l,\p_m)=0$ if $\p_l$ and $\p_m$ correspond to a {\it static} underlying manifold or the same manifold under rigid transformations;
    \item \textbf{Property 2}: $d(\p_l,\p_m)$ is small if $\p_l$ and $\p_m$ are similar, while being larger for more distinguishing patches. 
\end{itemize}

To meet the above properties, we define a manifold-to-manifold distance so as to measure the temporal distance between surface patches of irregular dynamic point cloud frames. We start from the {\it variation measure} on a manifold, and then derive its discrete counterpart on graphs, as demonstrated in Fig.~\ref{fig:manifold_point}.    

In particular, we consider Riemannian manifolds that are smooth and compact \cite{ting2010analysis}. 
To measure the variation on Riemannian manifolds, the Laplace–Beltrami operator is defined as the divergence of the gradient on Riemannian manifolds, which is a second-order differential operator. 
Let $\Delta_{\cM}$ denote the Laplace–Beltrami operator of a Riemannian manifold $\cM$. 
Given a smooth function $f$ on $\cM$, we have
\begin{equation}
    \Delta_{\cM}f = \text{div} (\nabla f),
\end{equation}
where div denotes the divergence operator and $\nabla f$ is the gradient of $f$. 

We propose to measure the distance between two manifolds $\cM_l$ and $\cM_m$ via the variation measurement.  
We consider normal coordinates $\n$ as one such smooth function on $\cM$, each of which is the coordinate of the normal $\n(x)$ in a neighborhood centered at a point $x$ on $\cM$. 
In our case, it suffices to consider normal coordinates as if the neighborhood is projected onto the tangent plane at
x. 
Then $\Delta_{\cM}\n$ measures the variation at each point by taking the second-order differential of normal coordinates on $\cM$, \ie, $\Delta_{\cM}\n = \text{div} (\nabla \n)$.
Further, we take the total variation of $\n$ as the variation measure of $\mathcal{M}$:   
\begin{equation}\label{eq:variation_manifold}
    V(\n,\mathcal{M}) \ = \ \frac{1}{|\cM|} \int_{\mathcal{M}} |\Delta_{\cM}\n(x)| d x. 
\end{equation}

Based on the variation measure $V(\n,\mathcal{M})$, we define the manifold-to-manifold distance between two manifolds $\cM_l$ and $\cM_m$ associated with normals $\n_l$ and $\n_m$ respectively as
\begin{equation}
    d(\cM_l,\cM_m)  \ = \ | V(\n_l,\cM_l) - V(\n_m,\cM_m)|,
    \label{eq:m2m_distance}
\end{equation}
which computes the difference between the total variation of $\cM_l$ and $\cM_m$. 

\subsection{Discrete Counterpart of the Manifold-to-Manifold Distance}

\begin{figure}[t]
  \begin{center}
    \begin{tabular}{c}
    \includegraphics[width=0.49\textwidth]{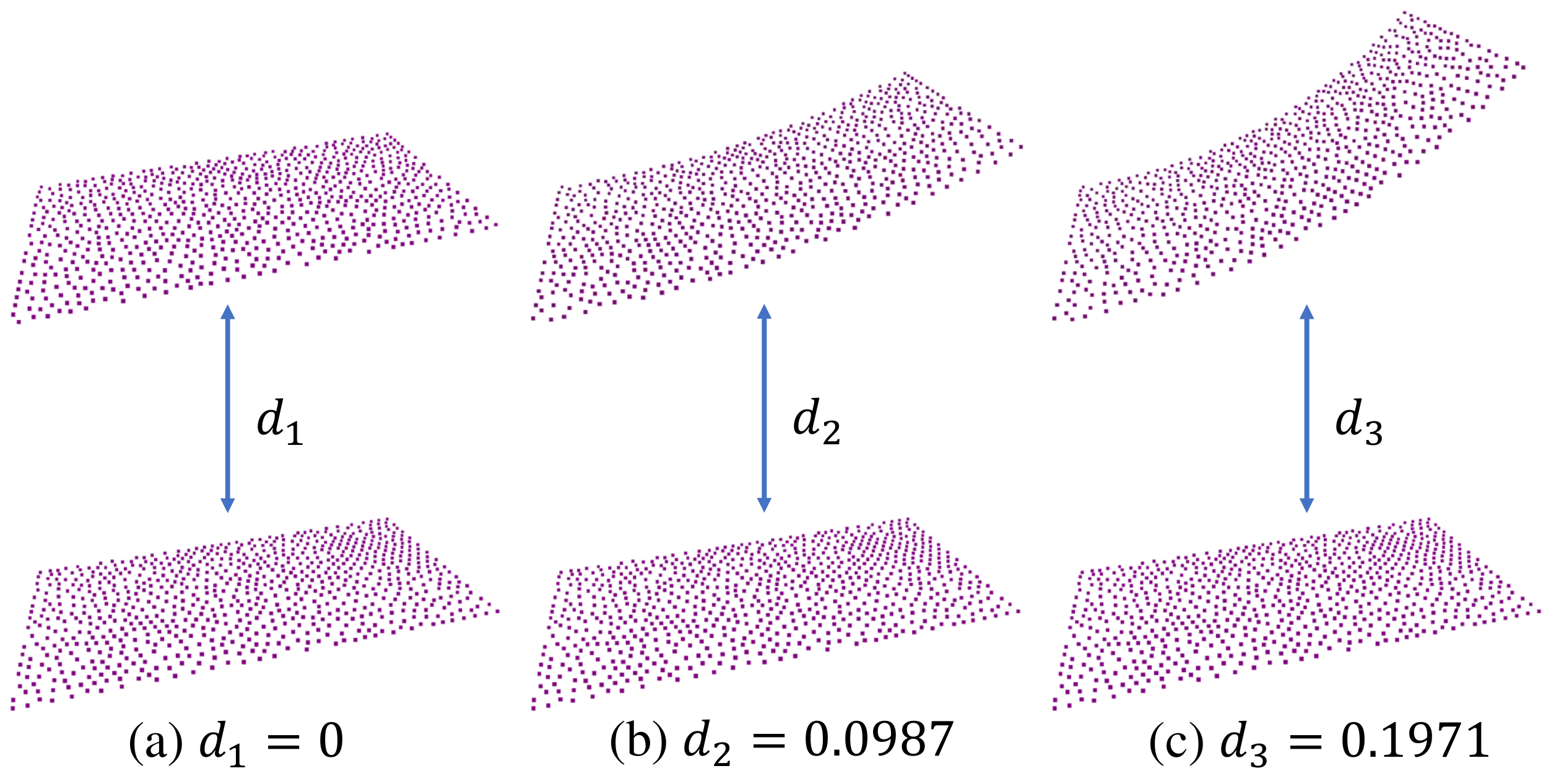}
  \end{tabular}
\end{center}
\vspace{-0.15in}
\caption{Examples of the distance measure between each pair of surface patches via the proposed discrete counterpart of Manifold-to-Manifold Distance in \eqref{eq:patch2patch_distance}.}
\label{fig:example_distance}
\end{figure}

Next, we derive the discrete counterpart of the defined manifold-to-manifold distance in \eqref{eq:m2m_distance}, which will be adopted to measure the distance between surface patches in adjacent point cloud frames. 

It has been proved that the graph Laplacian is a discrete approximation to the weighted Laplace-Beltrami operator \cite{ting2010analysis,hein2007graph}. 
In particular, as discussed in \cite{hein2007graph}, one may construct an $\epsilon$-neighborhood graph which can be seen as an approximation of the manifold. 
When the sample size $N$ goes to infinity and the neighborhood size $\epsilon$ goes to zero, the random walk graph Laplacian $\mathbf{L}_{\text{rw}} $ converges to the weighted Laplace-Beltrami operator on a manifold, \ie, 
\begin{equation}\label{eq:L_rw_approx}
\underset{\substack{N \rightarrow \infty\\ \epsilon \rightarrow 0}} {\lim}   \mathbf{L}_{\text{rw}} 
\sim \Delta_{\cM}.
\end{equation}

In the discrete domain, according to the definition of the random walk graph Laplacian, given a graph signal $\f \in \mathbb{R}^N$, the $i$-th element in $\L_{\text{rw}} \f$ is   
\begin{equation}
    [\L_{\text{rw}} \f]_i = \sum_{i \sim j} \frac{a_{i,j}}{d_{i,i}}(f_i - f_j),
    \label{eq:variation_point}
\end{equation}
where $i \sim j$ means node $i$ and node $j$ are connected. 
This computes the weighted variation of the graph signal on connected nodes, similar to the Laplace-Beltrami operator.  

Since a 3D point cloud is essentially discrete sampling from a Riemannian manifold $\cM$, we can operate the random walk graph Laplacian on the point cloud for variation measurement. 
Considering the normal coordinates $\n$ at points in the point cloud as the graph signal, we define the total variation of $\n$ as a discrete counterpart of \eqref{eq:variation_manifold}:
\begin{equation}\label{eq:variation_discrete}
    V(\n,\p) \ = \ \frac{1}{|\p|} \| \L_{\text{rw}} \n \|_1,
\end{equation}
where $|\p|$ is the number of points in the patch $\p$. \eqref{eq:variation_discrete} measures the structural variation in the surface patch of a point cloud. 

Similarly, we define a discrete counterpart of the manifold-to-manifold distance for two surface patches $\p_l$ and $\p_m$ associated with normals $\n_l$ and $\n_m$ respectively as
\begin{equation}
    d(\p_l,\p_m)  \ = | V(\n_l,\p_l) - V(\n_m,\p_m) |. 
    \label{eq:patch2patch_distance}
\end{equation}
This distance is permutation invariant and efficient to compute in a closed form.  


To demonstrate the effectiveness of the defined distance in \eqref{eq:patch2patch_distance}, we provide a few examples in Fig.~\ref{fig:example_distance} to show that the distance measurement satisfies both \textbf{Property 1} and \textbf{Property 2} in Section~\ref{subsec:m2m_distance}. 
As in Fig.~\ref{fig:example_distance}(a), the distance between the two surface patches is $0$ even though they are sampled in different irregular patterns from the same manifold, thus satisfying \textbf{Property 1}. 
When the two surface patches correspond to slightly different manifolds, the distance is nonzero but small as in Fig.~\ref{fig:example_distance}(b). 
It becomes larger for more distinguishing surface patches as in Fig.~\ref{fig:example_distance}(c), thus satisfying \textbf{Property 2}.

\subsection{Temporal Matching in Dynamic Point Clouds}
\label{subsec:temporal_matching}

Based on the defined temporal distance in \eqref{eq:patch2patch_distance}, we propose a temporal matching method so as to search temporally corresponding surface patches in adjacent point cloud frames. 

Given a target patch $\p_l$ in the current point cloud frame, we first build an unweighted $\epsilon$-neighborhood graph, where two points are connected with an edge weight $1$ if the Euclidean distance between them is smaller than a threshold $\epsilon$ and disconnected otherwise. 
Note that, the Euclidean distance is approximately the geodesic distance for neighboring points in the manifold underlying the surface patch. 
$\epsilon$ is assigned as the average of Euclidean distances between each pair of nearest points multiplied by a scalar $c$ (\eg, $c=5$).  

Then, we compute the random walk graph Laplacian $\L_{\text{rw}}$ from the constructed graph by definition, and operate it on the normal coordinates of points in $\p_l$ to acquire the variation measurement via \eqref{eq:variation_discrete}. 
The normal coordinate vector $\n_l^i \in \mathbb{R}^3$ at each point $i$ can be determined by fitting a local plane from neighboring points, via existing algorithms such as \cite{hoppe1992surface}.

\begin{figure*}
    \centering
    \includegraphics[width=0.8\linewidth]{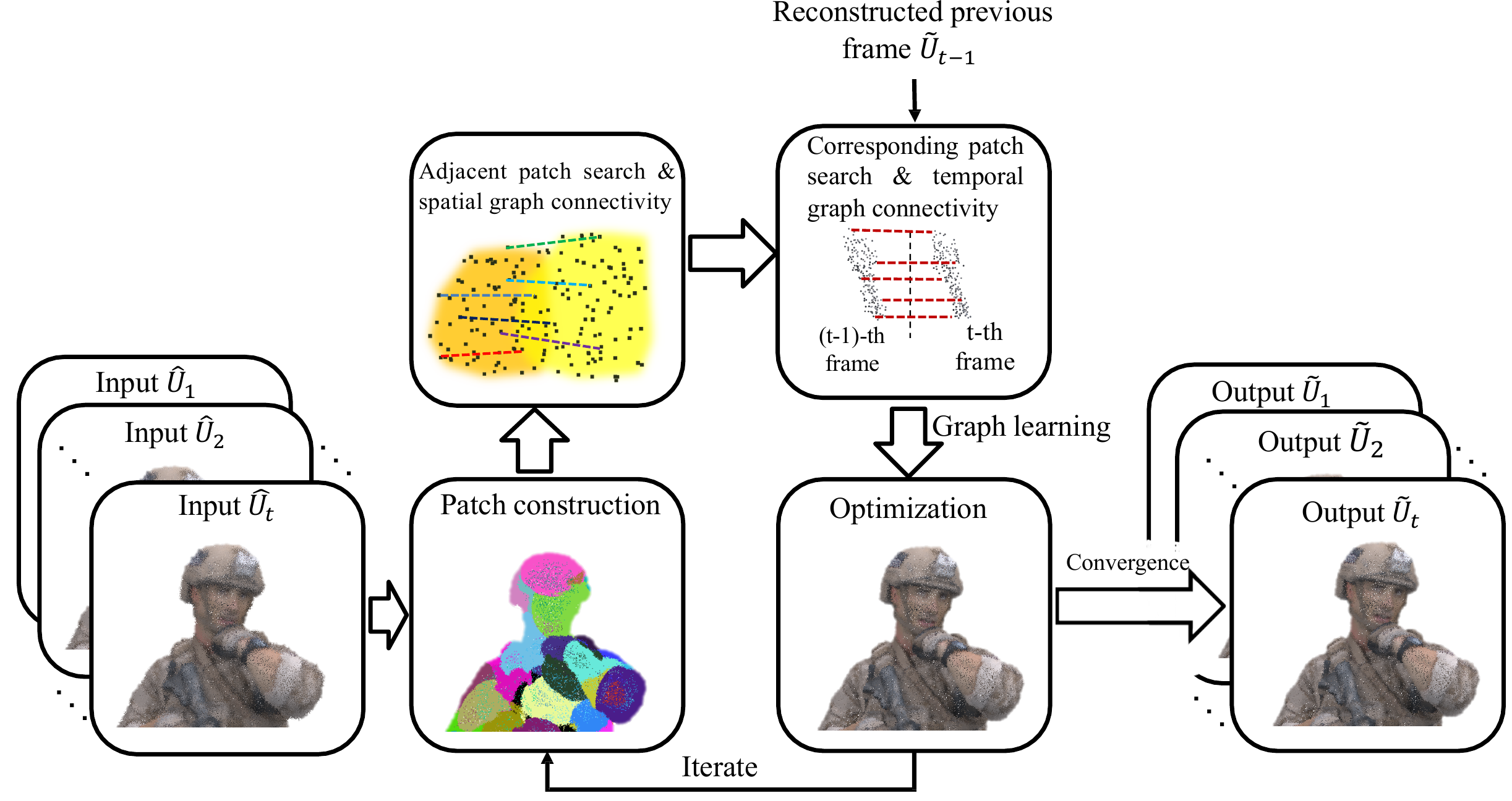}
    \centering
    \caption{The flowchart of the proposed dynamic point cloud denoising algorithm.}     
    \label{fig:flowchart}
\end{figure*}  

Next, we search the corresponding patch of $\p_l$ in the previous frame, which is the one with the smallest manifold-to-manifold distance in \eqref{eq:patch2patch_distance} to $\p_l$. 
That is, we follow the aforementioned graph construction and Laplacian calculation on reference patches to compute their variation measurement \eqref{eq:variation_discrete}, and find the one with the smallest manifold-to-manifold distance to $\p_l$.   
During the distance calculation in \eqref{eq:patch2patch_distance}, as $\n_l^i$ is of three dimensions, we compute \eqref{eq:patch2patch_distance} for the x-, y-, and z-dimension respectively, which leads to three distance measurements $[d_x, d_y, d_z]$. 
The final distance is the $l_2$ norm of this distance vector, \ie, 
\begin{equation}
    d(\p_l,\p_m)  \ = \  \sqrt{d_x^2 + d_y^2 + d_z^2}. 
\end{equation}

Further, in order to reduce the computation complexity, instead of global searching in the previous frame, we set a local search window that includes reference patches centering at the $\xi$-nearest neighbors of the collocated target patch center in terms of the Euclidean distance.



 \vspace{-0.05in}
\section{Problem Formulation}
\label{sec:method}
Leveraging on the proposed manifold-to-manifold distance, we propose a dynamic point cloud denoising algorithm to exploit the spatio-temporal correlation. 
A dynamic point cloud sequence $\mathcal{P}=\{\mathbf{U}_1,\mathbf{U}_2,...,\mathbf{U}_L\}$ consists of $L$ frames of point clouds. The coordinates $\mathbf{U}_t=[\mathbf{u}_{t,1}, \mathbf{u}_{t,2}, ... , \mathbf{u}_{t,N}]^\top\in \mathbb{R}^{N\times3}$ denote the position of each point in the point cloud in frame $t$, in which $\mathbf{u}_{t,i} \in \mathbb{R}^3$ represents the coordinates of the $i$-th point. 
Let $\mathbf{U}_t$ denote the ground truth coordinates of the $t$-th frame, and $\hat{\mathbf{U}}_{t-1}$, $\hat{\mathbf{U}}_t$ denote the noise-corrupted coordinates of the $(t-1)$-th and $t$-th frame respectively.
Given each noisy frame $\hat{\mathbf{U}}_t$, we aim to recover its underlying signal $\mathbf{U}_t$, exploiting the intra-frame correlation in $\hat{\mathbf{U}}_t$ as well as the inter-frame dependencies from the {\it reconstructed} previous frame $\widetilde{\mathbf{U}}_{t-1}$.  

As demonstrated in Fig.~\ref{fig:flowchart}, for a given dynamic point cloud, we perform denoising on each frame sequentially. The proposed algorithm consists of four major steps: 1) patch construction, where we form overlapped patches from chosen patch centers; 2) spatially nearest patch search and spatial graph construction, in which we search adjacent patches for each target patch in the current frame, and construct a spatial graph over these patches; 3) temporally corresponding patch search and temporal graph construction, in which we search the corresponding patch in the previous frame, and build inter-frame graph connectivities between corresponding patches; 4) optimization, where we formulate dynamic point cloud denoising as an optimization problem and solve it via the proposed algorithm. 
We perform step 2-4 iteratively during the optimization. 
Note that, the inter-frame reference is bypassed for denoising the first frame as there is no previous frame. 
Next, we discuss the four steps separately in detail. 

\subsection{Patch Construction}
We model both intra-frame and inter-frame dependencies on \textit{patch} basis. 
As mentioned in Section~\ref{subsec:m2m_distance}, we define a patch $\mathbf{p}_{t,l} \in \mathbb{R}^{(K+1) \times 3}$ in point cloud $\hat{\mathbf{U}}_t$ as a local point set of $K+1$ points, consisting of a centering point $\mathbf{c}_{t,l} \in \mathbb{R}^3$ and its $K$-nearest neighbors in terms of Euclidean distance. Then the entire set of patches in frame $t$ is     
\begin{equation}
    \mathbf{P}_t = \mathbf{S}_t \hat{\mathbf{U}}_t-\mathbf{C}_t,
    \label{eq:patch}
\end{equation}
where $\mathbf{S}_t\in\{0, 1\}^{(K+1)M\times N}$ is a sampling matrix to select points from point cloud $\hat{\mathbf{U}}_t$ so as to form $M$ patches of $(K+1)$ points per patch, and $\mathbf{C}_t \in \mathbb{R}^{(K+1)M\times 3}$ contains the coordinates of patch centers for each point.

Specifically, as each patch is formed around a patch center, we first select $M$ points from $\hat{\mathbf{U}}_t$ as the patch centers, denoted as $\{\mathbf{c}_{t, 1}, \mathbf{c}_{t, 2}, ... , \mathbf{c}_{t, M}\}\in \mathbb{R} ^ {M \times 3}\subset \hat{\mathbf{U}}_t$. In order to keep the patches distributed as uniformly as possible, we use the Farthest Point Sampling method \cite{fps}, \ie, we first choose a random point in $\hat{\mathbf{U}}_t$ as $\mathbf{c}_{t, 1}$, and add a point which holds the farthest distance to the previous patch centers as the next patch center, until there are $M$ points in the set of patch centers. We then search the $K$-nearest neighbors of each patch center in terms of Euclidean distance, which leads to $M$ patches in $\hat{\mathbf{U}}_t$.  

\subsection{Spatial Graph Connectivity and Initial Weights}
\label{section:Similar/Corresponding Patch Search}
Given each constructed patch in $\hat{\mathbf{U}}_t$, we search for its adjacent patches locally in $\hat{\mathbf{U}}_t$ so as to exploit the spatial correlation. 
Different from the temporal corresponding patch search in Section~\ref{subsec:temporal_matching} that finds patches with the same underlying surface in the observed 3D object, spatially adjacent patches correspond to different surfaces but with probably similar geometry. 

Given a target patch in the $t$-th frame $\hat{\mathbf{p}}_{t, l}, l\in \left[1, M\right]$, we seek its $K_s$ adjacent patches within the current frame $\hat{\mathbf U}_t$, denoted as $\{\hat{\mathbf{p}}_{t, m}\}_{m=1}^{K_s}$.
Specifically, we consider two patches as adjacent if their centers are $K_s$-nearest neighbors in terms of Euclidean distance.
Hence, we search the nearest $K_s$ patches of each patch as the neighbors based on the Euclidean distance between patch centers. 

Then, we construct a graph between $\hat{\mathbf{p}}_{t, l}$ and each of its adjacent patches $\hat{\mathbf{p}}_{t, m}$, leading to a $K_s$-nearest-patch graph on the entire point cloud. 
Each point in one patch is connected to its corresponding point in the other patch. 
For simplicity, we consider a pair
of points in adjacent patches as corresponding points if their coordinates relative to their respective centers are closest to each other. 
Due to patch overlaps, an edge may exist between two points within a patch if they are also corresponding points in two adjacent patches, as illustrated in Fig.~\ref{fig:graph}.

Further, we assign each pair of connected points $\{i,j\}$ with an initial edge weight $a_{i,j}$. 
Given a feature vector $\f_i$ associated with the $i$-th point, we adopt the commonly used Gaussian kernel to compute the edge weight from features, namely, 
\begin{align}
    a_{i,j} = \exp\left\{-(\mathbf{f}_i-\mathbf{f}_j)^{\top} (\mathbf{f}_i-\mathbf{f}_j) \right\}. 
\label{eq:initial_spatial_weight}
\end{align}
In particular, we employ two types of features: Cartesian coordinates and surface normals, where the normals are able to promote the piece-wise smoothness of the underlying manifold as discussed in \cite{dinesh18arxiv, Hu2020gsp}. 
Hence, we form a $6$-dimensional feature vector at each point $i$, \ie, $\f_i = [x_i,y_i,z_i,n_x^i,n_y^i,n_z^i]^{\top}$, where $[x_i,y_i,z_i]$ denotes the coordinates of point $i$, and $[n_x^i,n_y^i,n_z^i]$ denotes its normal vector.

The intra-frame connectivities are undirected and assigned with initial weights as in \eqref{eq:initial_spatial_weight}, which will be further optimized during dynamic point cloud denoising. 

\vspace{-0.1in}
\subsection{Temporal Graph Connectivity and Initial Weights}
\label{subsec:temporal_connectivity}
Given each target patch $\hat{\mathbf{p}}_{t, l}$ in $\hat{\mathbf{U}}_t$, we also search for its corresponding patch in $\widetilde{\mathbf{U}}_{t-1}$ so as to exploit the temporal relationship, which is the key to dynamic point cloud denoising. 
The proposed temporal matching has been described in Section~\ref{subsec:temporal_matching}, leading to 
the temporally corresponding patch $\widetilde{\mathbf{p}}_{t-1, m}$.

In order to leverage the inter-frame correlation and keep the temporal consistency, we connect corresponding patches between $\hat{\mathbf U}_t$ and $\widetilde{\mathbf U}_{t-1}$. 
Specifically, we connect points in each pair of corresponding patches if they have similar variation as measured by \eqref{eq:variation_point} and are close in coordinates relative to their respective centers, which considers both the local variation and relative location. 

\begin{figure}[t]
    \centering
    \includegraphics[width=\linewidth]{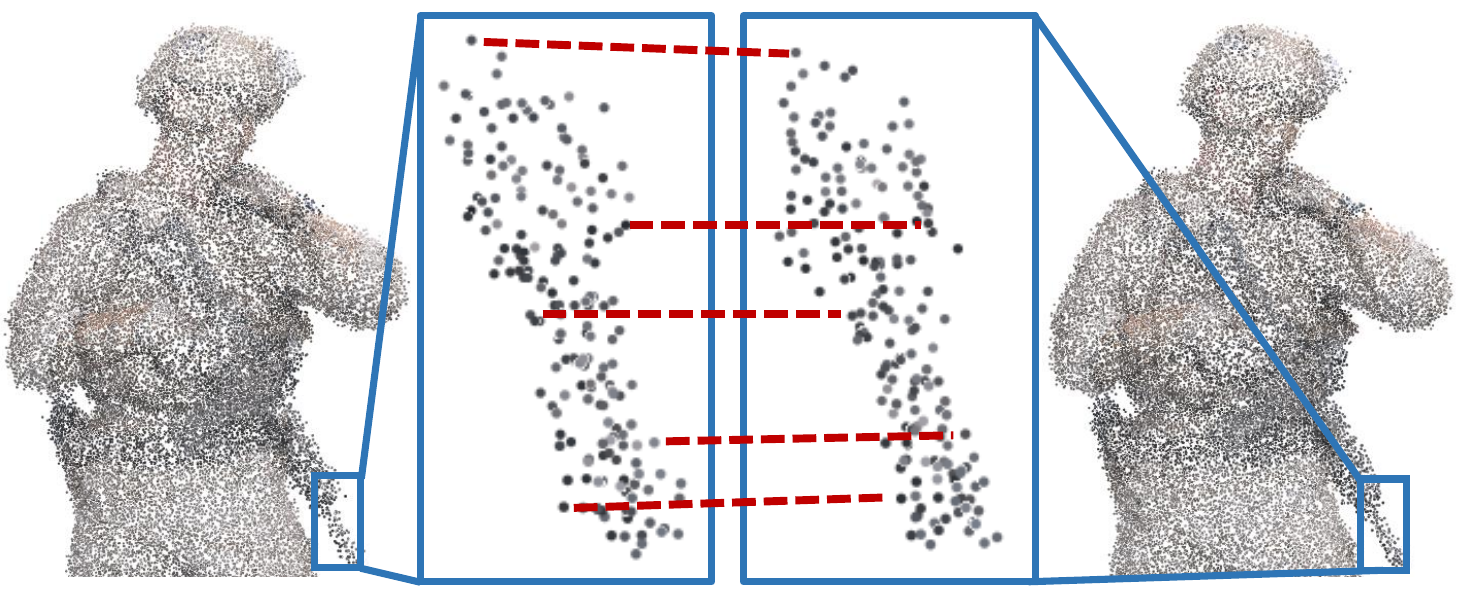}
    \vspace{-0.2in}
    \caption{Illustration of the graph connectivity between a pair of temporally corresponding patches.}
    \label{fig:graph}
\end{figure}

In particular, given the target patch $\hat{\mathbf{p}}_{t,l}$ and its corresponding patch $\widetilde{\mathbf{p}}_{t-1, m}$, we first compute the variation at each point as presented in \eqref{eq:variation_point}, based on the constructed graph and normals as in Section~\ref{subsec:temporal_matching}. 
Let $\L_{\text{rw},l}$ and $\L_{\text{rw},m}$ denote the random walk Laplacian constructed over $\hat{\mathbf{p}}_{t,l}$ and $\widetilde{\mathbf{p}}_{t-1, m}$ respectively, and let $\n_l$ and $\n_m$ denote the normals in the two patches respectively. 
For the $i$-th point in $\hat{\mathbf{p}}_{t,l}$ and the $j$-th point in $\widetilde{\mathbf{p}}_{t-1, m}$, we define the distance between the two points as
\begin{equation}
    d_{i,j} = \alpha \cdot \| [\L_{\text{rw},l} \n_l]_i - [\L_{\text{rw},m} \n_m]_j  \|_2^2 + (1-\alpha) \cdot \| \v_l^i - \v_m^j \|_2^2, 
    \label{eq:distance_point_temporal}
\end{equation}
where $[\cdot]_i$ denotes the $i$-th row in $\L_{\text{rw},l} \n_l$, which represents the variation around the $i$-th point. 
$\v_l^i$ and $\v_m^j$ denote the coordinates of the two points relative to their corresponding patch centers. 
$\alpha$ is a parameter to strike a balance between magnitudes of the affinities in the local variation and relative location. 

Based on \eqref{eq:distance_point_temporal}, we compute the distance between points in $\hat{\mathbf{p}}_{t,l}$ and $\widetilde{\mathbf{p}}_{t-1, m}$, and connect each point in $\hat{\mathbf{p}}_{t,l}$ to the point in $\widetilde{\mathbf{p}}_{t-1, m}$ with the smallest distance, as demonstrated in Fig.~\ref{fig:graph}. 
The initial edge weights of connected point pairs are all assigned as an exponential function of the manifold-to-manifold distance in \eqref{eq:patch2patch_distance}, \ie, 
\begin{equation}
    w_{l,m} = \exp\{-d(\hat{\mathbf{p}}_{t,l},\widetilde{\mathbf{p}}_{t-1, m})\}. 
    \label{eq:temporal_weight_init}
\end{equation}
While the edge weights in the spatial graph in 
\eqref{eq:initial_spatial_weight} are likely to be different for each pair of connected points in {\it adjacent} patches, the edge weight for all connected points in a pair of {\it temporally corresponding patches} is the {\it same}, which characterizes the temporal distance between the two corresponding patches.      
The initial temporal edge weight in \eqref{eq:temporal_weight_init} will also be further optimized during dynamic point cloud denoising.

As such, we construct spatio-temporal graphs over the patches, which serve as an initialization of the proposed dynamic point cloud denoising formulation.

\subsection{Formulation of Dynamic Point Cloud Denoising}

Assuming that dynamic point clouds lie on smooth manifolds both temporally and spatially, we formulate dynamic point cloud denoising as an optimization problem for each underlying point cloud frame $\mathbf{U}_t$, taking into account both the temporal consistency and intra-frame smoothness with respect to the constructed spatio-temporal graph that represents the underlying manifold. 

We first define the {\it temporal consistency} between two adjacent frames $\U_t$ and $\widetilde{\U}_{t-1}$ as the {\it temporal total variation} of corresponding patches $\{\mathbf{p}_{t,l},\widetilde{\mathbf{p}}_{t-1, m}\}$ in the two frames:   
\begin{equation}
\begin{split}
    \gF(\U_t,\widetilde{\U}_{t-1}) 
        & = \sum\limits_{l \sim m, l=1}^{M} w_{l,m} \| \mathbf{p}_{t,l} - \widetilde{\mathbf{p}}_{t-1, m} \|_2^2  \\
        & = \mathrm{tr}[(\P_t - \widetilde{\P}_{t-1})^{\top} \W_{t,t-1} (\P_t - \widetilde{\P}_{t-1})],
    \end{split}
    \label{eq:def_temporal_consistency}
\end{equation}
where $M$ is the number of patches in $\mathbf{U}_t$, $l \sim m$ represents $\mathbf{p}_{t,l}$ and $\widetilde{\mathbf{p}}_{t-1, m}$ are temporally corresponding patches, and $\W_{t,t-1} \in \mathbb{R}^{(K+1)M \times (K+1)M}$ is a diagonal matrix that encodes the temporal weights $w_{l,m}$ between temporally corresponding patches. 

Next, we formulate dynamic point cloud denoising as the joint optimization of the underlying point cloud $\mathbf{U}_t$ and the spatio-temporal graph representation encoded in the inter-frame weights $\W_{t,t-1}$ and the intra-frame graph Laplacian $\mathbf{L}_{t}$. 
Namely,we seek the optimal $\mathbf{U}_t$ and $\W_{t,t-1},\mathbf{L}_{t}$ to minimize an objective function including: 
1) a data fidelity term, which enforces $\mathbf{U}_t$ to be close to the observed noisy point cloud frame $\hat{\mathbf{U}}_t$; 
2) a temporal consistency term, which promotes the consistency between each patch in $\mathbf{P}_{t}$ in $\mathbf{U}_t$ and its correspondence in $\widetilde{\mathbf{P}}_{t-1}$ in the reconstructed previous frame $\mathbf{\widetilde{U}}_{t-1}$; 
3) a spatial smoothness term, which enforces smoothness of each patch in $\mathbf{U}_t$ with respect to the underlying graph encoded in the Laplacian $\mathbf{L}_{t} \in \mathbb{R}^{(K+1)M \times (K+1)M}$. 
The problem formulation is mathematically written as
\begin{equation} 
\begin{split}
& \min_{\mathbf{U}_t,\W_{t,t-1},\mathbf{L}_{t}} \ \ 
\begin{split}
&  \| \mathbf{U}_t - \hat{\mathbf{U}}_t \|_2^2  \\
& + \lambda_1 \cdot \mathrm{tr}[(\P_t - \widetilde{\P}_{t-1})^{\top} \W_{t,t-1} (\P_t - \widetilde{\P}_{t-1})] \\
& + \lambda_2 \cdot \mathrm{tr}(\P_t^\top \mathbf{L}_{t}\P_t) 
\end{split} \\ \\
& ~~~~~~\text{s.t.} ~~~~~~~~  \P_t = \mathbf{S}_{t} \mathbf{U}_{t} - \mathbf{C}_{t} \\
& ~~~~~~~~~~~~~~~~~~ 0 \leq [\mathbf{W}_{t, t-1}]_{i,i} \leq 1, \forall i \\
& ~~~~~~~~~~~~~~~~~~ [\mathbf{W}_{t, t-1}]_{i,j} = 0, \forall i \neq j
\end{split}
\label{eq:final}
\end{equation} 
$\lambda_1$ and $\lambda_2$ are weighting parameters for the trade-off among the data fidelity term, the temporal consistency term and the spatial smoothness term.
The trace $\mathrm{tr}(\cdot)$ is taken to compute the sum of the temporal consistency or the spatial smoothness in the x-, y- and z-coordinate. 
The first constraint is the definition of patches in \eqref{eq:patch}, while the other two constraints enforce $\mathbf{W}_{t, t-1}$ to be a diagonal matrix with diagonal entries as edge weights in the range of $[0,1]$. 

(\ref{eq:final}) is nontrivial to solve with three optimization variables. 
The variable $\W_{t,t-1}$ is dependent on $\P_t$ and $\widetilde{\P}_{t-1}$, while  
$\mathbf{L}_{t}$ is dependent on $\P_t$. 
In the next section, we develop an alternating minimization algorithm to reformulate and solve \eqref{eq:final}.

 \vspace{-0.05in}
\section{Algorithm Development}
\label{sec:algorithm}

We propose to address \eqref{eq:final} by alternately optimizing the underlying point cloud frame $\mathbf{U}_t$, the temporal weight matrix $\W_{t,t-1}$ and the intra-frame graph Laplacian $\L_t$. 
The iterations terminate when the difference in the objective between two consecutive iterations stops decreasing.

In particular, we first perform denoising on the first frame of a point cloud sequence by exploiting available intra-correlations ({\it i.e.}, $\lambda_1=0$). 
Then for each subsequent frame, we take advantage of the {\it reconstructed} previous frame as more robust reference than the noisy version, and alternately optimize the three optimization variables in \eqref{eq:final}. 

\subsection{Optimization of the point cloud $\mathbf{U}_t$}
In the first iteration, we initialize $\mathbf{W}_{t, t-1}$ and $\mathbf{L}_t$ from \eqref{eq:temporal_weight_init} and \eqref{eq:initial_spatial_weight} respectively. 
Then, we substitute the first constraint into the objective in \eqref{eq:final}, and set the derivative of the objective with respect to $\mathbf{U}_t$ to $0$. This leads to the closed-form solution of $\mathbf{U}_t$:
\begin{equation}
\begin{split}
& \left( \mathbf{I}+\lambda_1 \mathbf{S}_t^\top \mathbf{W}_{t,t-1}\mathbf{S}_t+\lambda_2 \mathbf{S}_t^\top \mathbf{L}_t \mathbf{S}_t\right) \mathbf{U}_t \\
= \ & \hat{\mathbf{U}}_t+ 
\lambda_1\mathbf{S}_t^\top \mathbf{W}_{t,t-1}(\mathbf{C}_t+\widetilde{\mathbf{P}}_{t-1}) + \lambda_2 \mathbf{S}_t^\top \mathbf{L}_t\mathbf{C}_t,
\label{eq:closed_form_U}
\end{split}
\end{equation} 
where $\mathbf{I} \in \mathbb{R}^{N \times N}$ is an identity matrix. 
\eqref{eq:closed_form_U} is a system of linear equations and thus can be solved efficiently.
Next, we employ the acquired solution of $\mathbf{U}_t$ to update $\mathbf{W}_{t, t-1}$ and $\mathbf{L}_t$ respectively. 

\vspace{-0.05in}
\subsection{Optimization of the Temporal Weight Matrix $\mathbf{W}_{t,t-1}$}
When we fix $\mathbf{L}_t$ as well as the updated $\mathbf{U}_t$, $\P_t$ is fixed accordingly and the optimization in \eqref{eq:final} becomes  
\begin{equation} 
\begin{split}
&\min_{\mathbf{W}_{t, t-1}} \mathrm{tr}[(\P_t - \widetilde{\P}_{t-1})^{\top} \W_{t,t-1} (\P_t - \widetilde{\P}_{t-1})] \\
& ~~\text{s.t.} ~~~ 0 \leq [\mathbf{W}_{t, t-1}]_{i,i} \leq 1, \forall i \\
& ~~~~~~~~~ [\mathbf{W}_{t, t-1}]_{i,j} = 0, \forall i \neq j
\end{split}
\label{eq:opt_temporal_weight}
\end{equation} 
This is essentially a linear program. 
Directly minimizing \eqref{eq:opt_temporal_weight} would lead to one pathological solution: $[\mathbf{W}_{t, t-1}]_{i,j}=0, \forall i,j$, as the objective is always non-negative according to \eqref{eq:def_temporal_consistency}. 
Topologically, this means patches in the current frame are all disconnected to the previous frame, defeating the goal of exploiting the temporal correlation.  
To avoid this solution, we constrain the trace of $\mathbf{W}_{t, t-1}$ to be larger than a
constant parameter $M'$, resulting in
\begin{equation} 
\begin{split}
&\min_{\mathbf{W}_{t, t-1}} \mathrm{tr}[(\P_t - \widetilde{\P}_{t-1})^{\top} \W_{t,t-1} (\P_t - \widetilde{\P}_{t-1})] \\
& ~~\text{s.t.} ~~~ 0 \leq [\mathbf{W}_{t, t-1}]_{i,i} \leq 1, \forall i \\
& ~~~~~~~~~ [\mathbf{W}_{t, t-1}]_{i,j} = 0, \forall i \neq j \\
& ~~~~~~~~~ \mathrm{tr}(\mathbf{W}_{t, t-1}) \geq M'
\end{split}
\label{eq:opt_temporal_weight_tr}
\end{equation} 

As $\mathbf{W}_{t, t-1}$ is a diagonal matrix in which we aim to optimize the diagonal, we rewrite the diagonal entries of $\mathbf{W}_{t, t-1}$ into a vector $\w \in \mathbb{R}^M$ as the optimization variable, which represents the temporal weights of $M$ pairs of temporally corresponding patches\footnote{Remind that connected points in a pair of corresponding patches share the same edge weight, as discussed in Section~\ref{subsec:temporal_connectivity}.}.  
In addition, let $\d  \in \mathbb{R}^{1 \times M}$ denote the squared difference between temporally corresponding patches, then we write \eqref{eq:opt_temporal_weight_tr} into the following form:  
\begin{equation} 
\begin{split}
&\min_{\w} ~~~ \w^{\top} \d \\
&~\text{s.t.} ~~~  0 \leq w_i \leq 1,  i=1,...,M \\
& ~~~~~~~~~ \sum_{i=1}^M w_i \geq M'
\end{split}
\label{eq:opt_temporal_weight_edit}
\end{equation} 

This is a linear program, and thus can be solved via off-the-shelf tools efficiently. 

\subsection{Optimization of the Intra-Frame Graph Laplacian $\mathbf{L}_t$}
Once we update the point cloud $\U_t$ and the temporal weight matrix $\mathbf{W}_{t, t-1}$, we keep them fixed and optimize the intra-frame graph Laplacian $\mathbf{L}_t$. 
The optimization problem in \eqref{eq:final} is now in the following form:
\begin{equation} 
\min_{\mathbf{L}_{t}} ~~~\text{tr}((\mathbf{S}_{t} \mathbf{U}_{t} - \mathbf{C}_{t})^\top \mathbf{L}_{t}(\mathbf{S}_{t} \mathbf{U}_{t} - \mathbf{C}_{t})). 
\label{eq:opt_Laplacian}
\end{equation} 

The objective in \eqref{eq:opt_Laplacian} can be rewritten as $\sum_{i \sim j} a_{i,j}\|\v_i-\v_j\|_2^2$, where $\v_i$ and $\v_j$ are the coordinates of connected points in the current frame $\U_t$, and $a_{i,j}$ is the edge weight between them.  
As in \cite{Hu19tsp}, we define the edge weight as 
\begin{align}
    a_{i,j} = \exp\left\{-(\mathbf{f}_i-\mathbf{f}_j)^{\top} \mathbf{M} (\mathbf{f}_i-\mathbf{f}_j) \right\}, 
\label{eq:weight}
\end{align}
where $\mathbf{f}_i \in \mathbb{R}^F$ and $\mathbf{f}_j \in \mathbb{R}^F$ are features associated with point $i$ and $j$ respectively. 
$\mathbf{M} \in \mathbb{R}^{F \times F} $ is the metric of the Mahalanobis distance \cite{mahalanobis1936} we aim to optimize, which captures both the relative importance of individual features and possible cross-correlation among features \cite{Hu19tsp}. 
$\mathbf{M}$ is required to be positive definite for a proper metric definition, {\it i.e.}, $\mathbf{M} \succ 0$. 

This leads to the following formulation:
\begin{equation}
\begin{split}
&\min_{\mathbf{M}}
    \sum_{i \sim j} \exp\left\{-(\mathbf{f}_i-\mathbf{f}_j)^{\top} \mathbf{M} (\mathbf{f}_i-\mathbf{f}_j) \right\} \, \|\v_i-\v_j\|_2^2\\
& \text{s.t.} \quad \,\mathbf{M} \succ 0; \;\;\;
\mathrm{tr}(\mathbf{M}) \leq C.
\label{eq:optimize_c_constraint}
\end{split}
\end{equation}
The constraint $\mathrm{tr}(\mathbf{M}) \leq C$ is to avoid the pathological solution $m_{i,i}=\infty, \forall i$ that results in edge weights $a_{i,j} = 0$, 
\ie, nodes in the graph are all isolated. 
With the additional trace constraint, we are able to construct a similarity graph to capture the intra-frame dependencies. 

As in Section~\ref{section:Similar/Corresponding Patch Search}, we consider both Cartesian coordinates and surface normals as the features associated with each point, forming a $6$-dimensional feature vector. 
Based on the constructed spatial graph connectivities discussed in Section~\ref{section:Similar/Corresponding Patch Search}, we then employ the feature difference at each pair of spatially connected points to learn the distance metric $\mathbf M$ in \eqref{eq:optimize_c_constraint} using the algorithm in \cite{Hu19tsp}. 
Substituting the optimized $\mathbf M$ into \eqref{eq:weight}, we acquire the edge weights and thus can compute the intra-frame graph Laplacian $\L_t$.  

A flowchart of the dynamic point cloud denoising algorithm is demonstrated in Fig.~\ref{fig:flowchart}, and an algorithmic summary is presented in Algorithm \ref{alg:Framwork}.

\vspace{-0.1in}

\begin{algorithm}[t]
  \caption{3D Dynamic Point Cloud Denoising}  
  \SetKwInOut{Input}{Input}\SetKwInOut{Output}{Output}
  \label{alg:Framwork}
    \Input{A noisy dynamic point cloud sequence $\hat{\mathcal{P}}=\{\hat{\mathbf{U}}_1,\hat{\mathbf{U}}_2,...,\hat{\mathbf{U}}_L\}$} 
    \Output{Denoised dynamic point cloud sequence  $\mathcal{P}=\{\mathbf{U}_1,\mathbf{U}_2,...,\mathbf{U}_L\}$} 
    \For{$\hat{\mathbf{U}}_t$ in $\hat{\mathcal{P}}$}{

    \Repeat{convergence}{
        Select $M$ points (set $\mathbf{C}_t$) as patch centers; \\
        \For{$\mathbf{c}_l$ in $\mathbf{C}_t$}{
            Find $K$-nearest neighbors of $\mathbf{c}_l$; \\
            Build patch $\hat{\mathbf{p}}_{t,l}$; \\
            Add $\hat{\mathbf{p}}_{t,l}$ to $\hat{\mathbf{P}}_t$; \\
            Find $\xi$-nearest patches of $\hat{\mathbf{p}}_{t,l}$ in the previous frame $\widetilde{\U}_{t-1}$;\\
            Calculate the distances between the $\xi$ patches and $\hat{\mathbf{p}}_{t,l}$ using \eqref{eq:patch2patch_distance};\\
            Select the patch with the smallest distance as $\tilde{\mathbf{p}}_{t-1,m}$;\\
            \For {i in $\hat{\mathbf{p}}_{t,l}$}{
                Find a node $j$ in $\tilde{\mathbf{p}}_{t-1,m}$ with the smallest distance calculated by \eqref{eq:distance_point_temporal};
            }
            Adjust the order of nodes in patch $\tilde{\mathbf{p}}_{t-1,m}$ to keep correspondence;\\
            Add $\tilde{\mathbf{p}}_{t-1,m}$ to $\tilde{\mathbf{P}}_{t-1}$;\\
        }
        \For{$\hat{\mathbf{p}}_{t,l}$ in $\hat{\mathbf{P}}_t$}{
            $\mathcal{B}_l$ $\leftarrow$ the nearest $K_s$ patches to $\hat{\mathbf{p}}_{t,l}$;\\
            Find corresponding node pairs from $\mathcal{B}_l$ via the algorithm in Section~\ref{section:Similar/Corresponding Patch Search}\\
        }
        
        \eIf{this is the first iteration}{
            Initialize $\mathbf{W}_{t, t-1}$ and $\mathbf{L}_t$ from \eqref{eq:temporal_weight_init} and \eqref{eq:initial_spatial_weight} \\
        }
        {
            Solve \eqref{eq:opt_temporal_weight_edit} to update $\mathbf{W}_{t, t-1}$; \\  
            Solve \eqref{eq:optimize_c_constraint} to update $\mathbf{L}_t$;  \\
        }
        Solve \eqref{eq:closed_form_U} to update $\hat{\mathbf{U}}_t$; \\

        }
        The denoised result $\widetilde{\mathbf{U}}_t$ serves as the input for the denoising of the next frame.
    }
\end{algorithm}  

\section{Experimental Results}
\label{sec:results}
\begin{table*}[htbp]
\centering
\caption{MSE comparison among different methods with Gaussian noise.}
\begin{tabularx}{0.9\textwidth}{|c|Y|Y|Y|Y|Y|Y|Y|}
\hline
\textbf{Model} & \textbf{Noisy} & \textbf{RIMLS} & \textbf{MRPCA} & \textbf{NLD} & \textbf{LR} & \textbf{FGL} & \textbf{Ours} \\ \hline \hline
\multicolumn{8}{|c|}{$\sigma=0.1$} \\ \hline
Soldier & 3.0109  & 2.5549 & 2.2476 & 2.4152 & 2.1813 & 2.4856 & $\mathbf{2.1521}$ \\ \hline
Longdress & 2.9843 & 2.5463 & 2.2288 & 2.3865 & 2.1507 & 2.4230 & $\mathbf{2.1326}$ \\ \hline
Loot & 2.9875  & 2.4407 & 2.1728 & 2.3783 & 2.0936 & 2.3744 & $\mathbf{2.0877}$ \\ \hline
Redandblack & 2.9597  & 2.5615 & 2.2496 & 2.3901 & 2.1640 & 2.4968 & $\mathbf{2.1540}$ \\ \hline
Andrew & 1.5606  & 1.3676 & 1.1210 & 1.3473  & 1.1326  & 1.1895 & $\mathbf{1.1087 }$ \\ \hline
David & 1.6652  & 1.3232 & 1.1196 & 1.4193  & 1.1400  & 1.1427 & $\mathbf{1.0842 }$ \\ \hline
Phil & 1.5579  & 1.4578 & 1.1320 & 1.3288  & 1.1408  & 1.2250 & $\mathbf{1.1284 }$ \\ \hline
Ricardo & 1.6629  & 1.4438 & 1.1470 & 1.4250  &  1.1658  & 1.2118 & $\mathbf{1.1438  }$ \\ \hline
Sarah & 1.5744  & 1.2536 & 1.0828 & 1.3284  & 1.1022  & 1.1393 & $\mathbf{1.0683  }$ \\ \hline 
Average & 2.2182 & 1.8833 & 1.6112 & 1.8243 & 1.5857 & 1.7431 & $\mathbf{1.5622 }$ \\ \hline 
\hline
\multicolumn{8}{|c|}{$\sigma=0.2$} \\ \hline
Soldier & 4.4096  & 4.1654 & 2.7633 & 3.8865 & 2.7932 & 3.0822 & $\mathbf{2.6482}$ \\ \hline
Longdress & 4.3628  & 3.7721 & 2.6474 & 3.8328 & 2.7156 & 2.9930 & $\mathbf{2.5939}$ \\ \hline
Loot & 4.3991 & 4.1796  & 2.5685 & 3.8974 & 2.6430 & 2.8973 
& $\mathbf{2.4626}$ \\ \hline
Redandblack & 4.3082  & 3.9882 & 2.6963 & 3.8373 & 2.7586 & 3.1277 & $\mathbf{2.6715}$ \\ \hline
Andrew & 2.3655  & 2.2613 & 1.8468 & 2.2303 & 1.8086 & 1.4190 & $\mathbf{1.3010   }$ \\ \hline
David & 2.5969  &  2.4020 & 2.0359 & 2.4571 & 2.0172 & 1.3731 & $\mathbf{1.2695  }$ \\ \hline
Phil & 2.3470  & 2.3984 & 1.7892 & 2.1940 & 1.7615 & 1.4961 & $\mathbf{1.3508  }$ \\ \hline
Ricardo & 2.5907  & 2.6432 & 2.0331 & 2.4472 & 2.0362 & 1.5022 & $\mathbf{1.3740 }$ \\ \hline
Sarah & 2.4083  & 2.1961 & 1.8006 & 2.2491 & 1.7739 & 1.3441 &  $\mathbf{1.2206 }$ \\ \hline 
Average & 3.3098  & 3.0519 & 2.2423 & 3.0035 & 2.2564 & 2.1372 & $\mathbf{1.8769}$ \\ \hline 
\hline
\multicolumn{8}{|c|}{$\sigma=0.3$} \\ \hline
Soldier & 5.6965  & 5.7052 & 4.4023 & 5.4099 & 4.1006 & 3.6666 & $\mathbf{3.1436}$ \\ \hline
Longdress & 5.6295  & 5.7341 & 4.2085 & 5.3498 & 3.9991 & 3.5898 & $\mathbf{3.1330}$ \\ \hline
Loot & 5.7190  & 5.8351 & 4.1514 & 5.3752 & 3.9383 & 3.4449 
& $\mathbf{2.8710}$ \\ \hline
Redandblack & 5.5436  & 5.5522 & 4.2195 & 5.2830 & 4.0068 & 3.7656 & $\mathbf{3.2671}$ \\ \hline
Andrew & 3.1835  & 3.2359 & 2.8769 & 3.1033 & 2.8901 & 1.6852 & $\mathbf{1.5415  }$ \\ \hline
David & 3.5441  & 3.5993 & 3.2316 & 3.4541 & 3.2668 & 1.6674 & $\mathbf{1.5228   }$ \\ \hline
Phil & 3.1420  & 3.1923 & 2.7988 & 3.0541 & 2.7889 & 1.8257 & $\mathbf{1.7739  }$ \\ \hline
Ricardo & 3.5292  &  3.5843 & 3.1989 & 3.4363 & 3.2460 & 1.9207 & $\mathbf{1.7240 }$ \\ \hline
Sarah & 3.2537  & 3.3048 & 2.8941 & 3.1657 & 2.9157 & 1.6013 & $\mathbf{1.4262  }$ \\ \hline 
Average & 4.3608  & 4.4159 & 3.5536 & 4.1813 & 3.4614 & 2.5741 & $\mathbf{2.3045}$ \\ \hline 
\hline
\multicolumn{8}{|c|}{$\sigma=0.4$} \\ \hline
Soldier & 6.9185  & 7.0302 & 6.1384 & 6.7725 & 5.7047 & 4.3620 & $\mathbf{3.8756}$ \\ \hline
Longdress & 6.8394  & 6.9530 & 5.9919 & 6.6840 & 5.6054 & 4.2953 & $\mathbf{3.8573}$ \\ \hline
Loot & 7.0044  & 7.1230 & 6.0280 & 6.8178 & 5.5869 & 4.0729 
& $\mathbf{3.4782}$ \\ \hline
Redandblack & 6.6865  & 6.7996 & 5.8635 & 6.5250 & 5.4822 & 4.3422 & $\mathbf{4.0019}$ \\ \hline
Andrew & 4.0003  & 4.0538 & 3.8059 & 3.9498 & 3.8873 & 2.0138 & $\mathbf{1.8995}$ \\ \hline
David & 4.4872  & 4.5381 & 4.2870 & 4.4326 & 4.3771 & 2.1608 & $\mathbf{1.9799}$ \\ \hline
Phil & 3.9334  & 3.9895 & 3.7231 & 3.8845 & 3.7688 & 2.2620 & $\mathbf{2.0676}$ \\ \hline
Ricardo & 4.4677  & 4.5249 & 4.2508 & 4.4117 & 4.3425 & 2.5251 & $\mathbf{2.2690 }$ \\ \hline
Sarah & 4.1130  & 4.1694 & 3.8849 & 4.0466 & 3.9581 & 1.9230 & $\mathbf{1.7478}$ \\ \hline 
Average & 5.3834  & 5.4646 & 4.8859 & 5.2805 & 4.7459 & 3.1063 & $\mathbf{2.7974}$ \\ \hline 

\end{tabularx}
\label{tb:mse}
\end{table*}

\begin{table*}[htbp]
\centering
\caption{GPSNR comparison among different methods with Gaussian noise (unit: dB).}
\begin{tabularx}{0.9\textwidth}{|c|Y|Y|Y|Y|Y|Y|Y|}
\hline
\textbf{Model} & \textbf{Noisy} & \textbf{RIMLS} & \textbf{MRPCA} & \textbf{NLD} & \textbf{LR} & \textbf{FGL} & \textbf{Ours} \\ \hline \hline
\multicolumn{8}{|c|}{$\sigma=0.1$} \\ \hline
Soldier & 13.8011  & 17.3398 & 19.3423 & 18.6466 & 19.8049  & 17.3346 & $\mathbf{19.9349}$ \\ \hline
Longdress & 13.7349  & 16.9798 & 19.1825 & 18.6686 & $\mathbf{19.8086}$ & 17.6164 & 19.8027 \\ \hline
Loot & 14.3811  & 18.1853 & 20.3898 & 19.5237 & $\mathbf{21.2139}$ & 18.7960 & 21.0921 \\ \hline
Redandblack & 14.9472  & 18.1420 & 20.3042 & 19.7765 & $\mathbf{20.8990}$ & 18.0948 & 20.7952 \\ \hline
Andrew & 23.7454  & 27.0087 & 29.6703 & 27.4610 & $\mathbf{29.9695}$  & 29.2948 & 29.7194  \\ \hline
David & 20.7492  & 25.3825 & 28.0130 & 24.6030 & 27.7643 & 27.5613 & $\mathbf{27.8535  }$ \\ \hline
Phil & 29.4216  & 31.6705 & 35.1765 & 33.2929 & $\mathbf{35.3307 }$ & 34.1603 & 34.9090   \\ \hline
Ricardo & 25.8886  & 29.4784 & 32.6949 & 29.5197 & 32.2186 & 31.8458 & $\mathbf{32.4774 }$   \\ \hline
Sarah & 19.3478  & 23.8216 & 25.9225 & 23.4533 & $\mathbf{26.1747 }$ & 25.4024  & 25.7537  \\ \hline 
Average & 19.5574  & 23.1121 & 25.6329 & 23.8828 & $\mathbf{25.9094}$ & 24.4563 & 25.8153 \\ \hline 
\hline
\multicolumn{8}{|c|}{$\sigma=0.2$} \\ \hline
Soldier & 8.3618  & 11.1889 & 15.6370 & 11.2064 & 15.2246 & 13.3409 & $\mathbf{16.2853}$ \\ \hline
Longdress & 8.2656  & 11.2579 & 16.1272 & 11.1816 & 15.4283 & 13.4528 & $\mathbf{16.2650}$ \\ \hline
Loot & 8.8221  & 12.6842 & 17.3280 & 11.6637 & 16.7246 & 14.7706 & $\mathbf{17.9732}$ \\ \hline
Redandblack & 9.5163  & 12.1459 & 16.9286 & 12.2156 & 16.2573 & 13.9137 & $\mathbf{16.9523}$ \\ \hline
Andrew & 17.9095  & 20.7434 & 22.2928 & 19.8357 & 22.3684 & 26.9396 & $\mathbf{27.6431 }$ \\ \hline
David & 14.7879  & 17.3617 & 18.9538 & 16.5962 & 18.9736 & 25.0981  & $\mathbf{25.9291}$   \\ \hline
Phil & 23.7088  & 26.2104 & 28.3297 & 25.7554 & 28.3826 & 31.5465 & $\mathbf{32.7310 }$ \\ \hline
Ricardo & 19.9857  & 19.9465 & 24.0801 & 21.7578 & 23.9155  & 28.8548  & $\mathbf{30.0770  }$ \\ \hline
Sarah & 13.4032  & 16.4626 & 18.2512 & 15.4942 & 18.3083 & 23.4074  & $\mathbf{24.0133 }$ \\ \hline 
Average & 13.8623  & 16.4446 & 19.7698 & 16.1896 & 19.5092 & 21.2583 & $\mathbf{23.0966}$ \\ \hline 
\hline
\multicolumn{8}{|c|}{$\sigma=0.3$} \\ \hline
Soldier & 5.1182  & 5.1179 & 9.1878 & 6.8187 & 9.6504 & 10.9496 & $\mathbf{13.5838}$ \\ \hline
Longdress & 5.0226  & 4.9900 & 9.5114 & 6.7567 & 9.7090 & 10.9036 & $\mathbf{13.2192}$ \\ \hline
Loot & 5.4727  & 5.4283 & 10.2855 & 7.3720 & 10.6216 & 12.2262 & $\mathbf{15.4670}$ \\ \hline
Redandblack & 6.2721  & 6.2723 & 10.4824 & 7.9642 & 10.7746 & 11.2215 & $\mathbf{13.5975}$ \\ \hline
Andrew & 14.3132  & 14.3123 & 16.6570 & 15.5086 & 16.6501 & 24.7183 & $\mathbf{25.7365  }$ \\ \hline
David & 11.1894  & 12.7832 & 13.4009 & 12.2866 & 13.2955 & 22.3039  & $\mathbf{23.9304 }$ \\ \hline
Phil & 20.1802  & 20.1435 & 22.6800 & 21.4216 & 22.7227 & 22.6199 & $\mathbf{26.5766}$   \\ \hline
Ricardo & 16.4341  & 16.4004 & 18.6651 & 17.5236 & 18.4746 & 25.0456 & $\mathbf{26.9095  }$ \\ \hline
Sarah & 9.7971  & 9.7595 & 12.3346 & 11.0410 & 12.2891 & 20.9865 & $\mathbf{22.3066 }$ \\ \hline 
Average & 10.4222  & 10.5786 & 13.6894 & 11.8548 & 13.7986 & 17.8861 & $\mathbf{20.1475}$ \\ \hline 
\hline
\multicolumn{8}{|c|}{$\sigma=0.4$} \\ \hline
Soldier & 2.7953  & 2.7613 & 5.3401  & 3.9304 & 5.7206 & 8.2346 & $\mathbf{10.4425}$ \\ \hline
Longdress & 2.6923  & 2.6577 & 5.4387 & 3.8577 & 5.6411 & 8.2144 & $\mathbf{10.1229}$ \\ \hline
Loot & 3.0694  & 3.0324 & 6.0255 & 4.3101 & 6.3995 & 9.3759 & $\mathbf{12.4929}$ \\ \hline
Redandblack & 4.0372  & 4.0018 & 6.7228 & 5.2440 & 7.0241 & 9.1485 & $\mathbf{10.5798}$ \\ \hline
Andrew & 11.8127  & 11.7804 & 13.3392 & 12.6197 & 13.1625 & 22.3828  & $\mathbf{23.6789 }$ \\ \hline
David & 8.6440  & 8.6150 & 10.0778 & 9.4079 & 9.8966 & 18.6479  & $\mathbf{21.0552}$ \\ \hline
Phil & 17.6925  & 17.6634 & 19.3264 & 18.5298 & 19.2215 & 25.7831 & $\mathbf{27.2161 }$ \\ \hline
Ricardo & 13.9319  & 13.9021 & 15.4137 & 14.6882 & 15.1792 & 21.1708 & $\mathbf{22.9000 }$  \\ \hline
Sarah & 7.2136  & 7.1765 & 8.8312 & 8.0494 & 8.6941 & 18.5834  & $\mathbf{20.0646 }$\\ \hline 
Average & 7.9877  & 7.9545 & 10.0573 & 8.9587 & 10.1044 & 15.7268 & $\mathbf{17.6170}$ \\ \hline 

\end{tabularx}
\label{tb:gpsnr}
\end{table*}

\vspace{-0.05in}
\subsection{Experimental Setup}
\label{subsec:setup}
We evaluate our algorithm by testing on two benchmarks, including four MPEG sequences (\textit{Longdress}, \textit{Loot}, \textit{Redandblack} and \textit{Soldier}) from \cite{MPEG} and five MSR sequences (\textit{Andrew}, \textit{David}, \textit{Phil}, \textit{Ricardo} and \textit{Sarah}) from \cite{loop2016microsoft}.
We randomly choose six consecutive frames in each sequence as the sample data: frame $601$-$606$ in \textit{Soldier}, frame $1201$-$1206$ in \textit{Loot}, frame $1201$-$1206$ in \textit{Longdress}, frame $1501$-$1506$ in \textit{Redandblack}, frame $61$-$66$ in \textit{Andrew}, frame $61$-$66$ in \textit{David}, frame $61$-$66$ in \textit{Phil}, frame $71$-$76$ in \textit{Ricardo} and frame $61$-$66$ in \textit{Sarah}.
We perform down-sampling with the sampling rate of $0.05$ prior to denoising since the number of points in each frame is about $1$ million. 

We test on two types of noise:
1) synthetic white Gaussian noise with a range of variance $\sigma=\{0.1,0.2,0.3,0.4\}$ added to the clean point clouds in the datasets for objective comparison; 
2) simulated LiDAR noise to mimic the real-world LiDAR noise. Due to the lack of real-world noisy dynamic point cloud datasets that are publicly available, we adopt a LiDAR simulator---the simulation package Blensor \cite{gschwandtner2011blensor} to produce realistic noise in point clouds. We use Velodyne HDL-64E2 as the scanner model in simulations and set the noise level to 0.01.

Due to the lack of previous dynamic point cloud denoising approaches, we compare our algorithm with five competitive static point cloud denoising methods: RIMLS \cite{A2009Feature}, MRPCA \cite{Mattei2017Point}, NLD \cite{deschaud10}, LR \cite{sarkar2018structured} and FGL \cite{Hu19tsp}. 
We perform each static denoising method frame by frame independently on dynamic point clouds. 
Among them, FGL \cite{Hu19tsp} is our baseline method, which we extend to dynamic point cloud denoising by introducing the temporal consistency. 
When the weighting parameter $\lambda_1$ in \eqref{eq:final} for the temporal consistency term is assigned $0$, our method defaults to FGL. 

We employ two evaluation metrics to measure the objective quality of the reconstructed point clouds: 
1) a point-to-point metric: Mean Squared Error (MSE), which directly measures the error between points in the reconstructed point cloud and the ground truth;
2) a point-to-plane metric: Geometric PSNR (GPSNR) \cite{Tian17}, which measures projected error vectors along normal directions for better surface structural description. In our experiments, the peak value in GPSNR is set to $5$.  
The lower MSE and the higher GPSNR is, the smaller the difference between two point clouds is.

The parameter settings are as follows. 
We divide each point cloud into $M=0.5N$ patches, where $N$ is the number of points in the point cloud. 
The number of nearest neighbors for each patch center to form a patch is $K=30$, and the number of nearest patches of each target patch in the previous frame is $\xi=10$.
Each patch is connected with $K_s=10$ most similar patches spatially.
Besides, given the first frame in each dataset, we set $\lambda_1=0$ as there is no previous frame as reference.
We assign the lower bound $M'$ in \eqref{eq:opt_temporal_weight_edit} as $M'=0.9M$, and the upper bound $C$ of the trace of $\mathbf M$ in \eqref{eq:optimize_c_constraint} as $5$.

\vspace{-0.05in}
\subsection{Experimental Results}
\subsubsection{Objective Results} 
We list the denoising results of comparison methods measured in MSE and GPSNR in Tab.~\ref{tb:mse} and Tab.~\ref{tb:gpsnr} respectively, and mark the lowest MSE and the highest GPSNR in bold. 
Our method outperforms all the five static point cloud denoising approaches on the nine datasets in general, especially at higher noise levels. 

Specifically, we achieve reduction of the MSE by $42.35\%$ on average over RIMLS, $30.52\%$ on average over MRPCA, $40.23\%$ on average over NLD, $29.12\%$ on average over LR, and $10.67\%$ on average over FGL. 
In terms of GPSNR, we achieve performance gain by $7.15$ dB on average over RIMLS, $4.38$ dB on average over MRPCA, $6.45$ dB on average over NLD, $4.34$ dB on average over LR, and $1.84$ dB on average over FGL. 
This validates the effectiveness of the learned temporal consistency.  
Further, we observe that we achieve larger gain at challenging higher noise levels.   
This gives credits to the proposed temporal consistency with respect to the underlying manifold based on the  manifold-to-manifold distance. 


Also, we achieve larger gain over dynamic point clouds with {\it slower motion} due to the stronger temporal correlation. 
For instance, the average MSE reduction over the comparatively static \textit{Sarah} is $51.87\%$, while that over \textit{Redandblack} is $37.97\%$ with more dynamic motion in the tested frames.



\subsubsection{Subjective Results} 
\begin{figure*}[htbp]
    \centering
    \includegraphics[width=\linewidth]{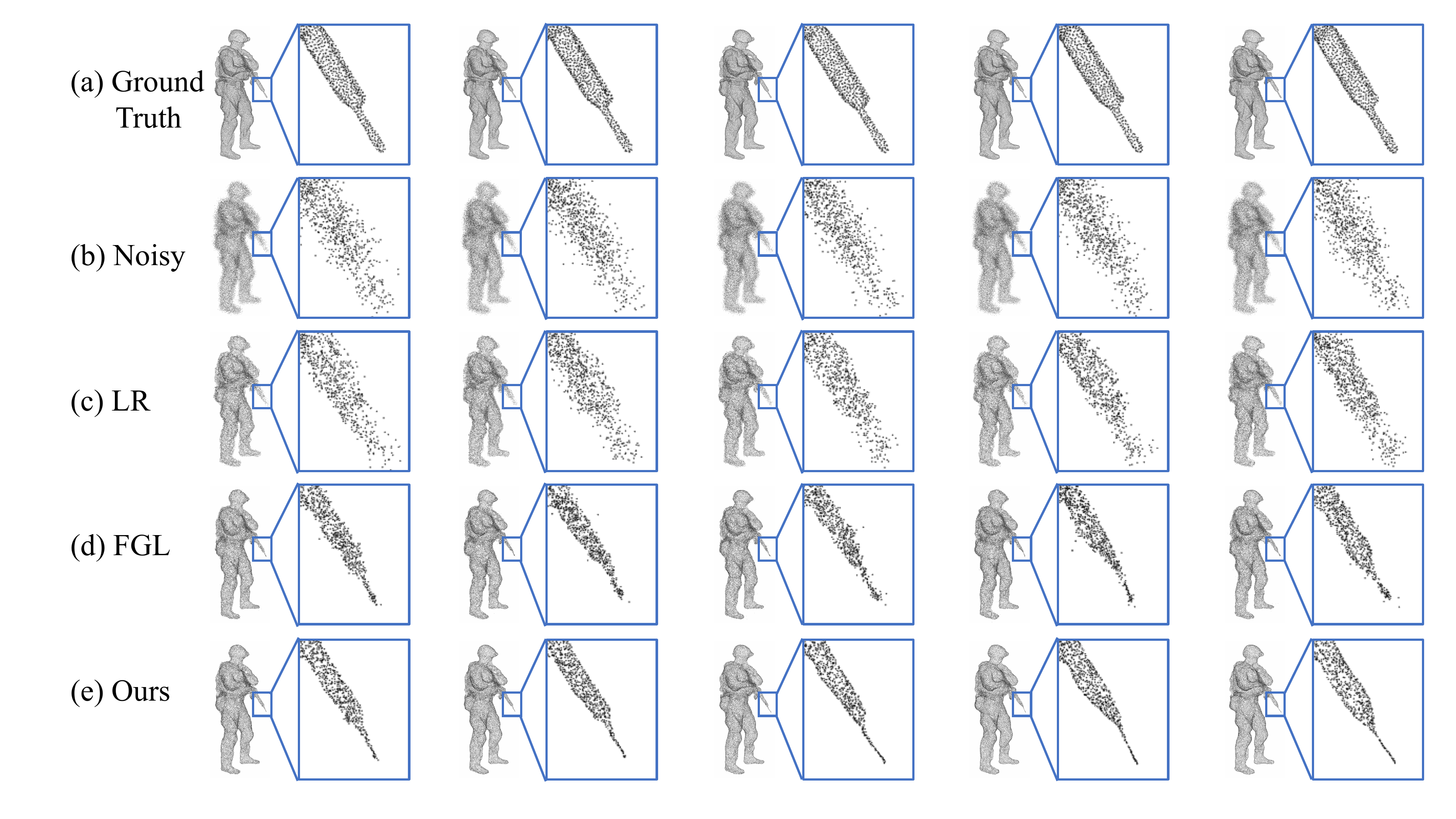}
    \caption{Comparison results from Gaussian noise ($\sigma=0.3$) for \textit{Soldier}: (a) The ground truth; (b) The noisy point cloud; (c) The denoised result by LR; (d) The denoised result by FGL; (e) The denoised result by our algorithm. Colors are not shown for clear demonstration of geometry denoising.}
    \label{fig:soldier}
\end{figure*}

As illustrated in Fig.~\ref{fig:soldier}, the proposed method also improves the visual results significantly, especially in local details and temporal consistency. 
We choose $5$ consecutive frames in $\textit{Soldier}$ under the noise variance $\sigma=0.3$, and show the visual comparison with LR and FGL because they are the nearest two competitors to our method in the objective performance. 
We see that, the shape of the gun is not well reconstructed in the results of LR, and both the results of LR and FGL still suffer from noise and even outliers. 
In contrast, our results preserve the local structure of the gun much better, and mitigate the outliers significantly. 
Further, our results are more consistent in the temporal domain. 

\subsubsection{Results on Simulated Real-World Noise}
\begin{figure*}[htbp]
    \centering
    \includegraphics[width=\linewidth]{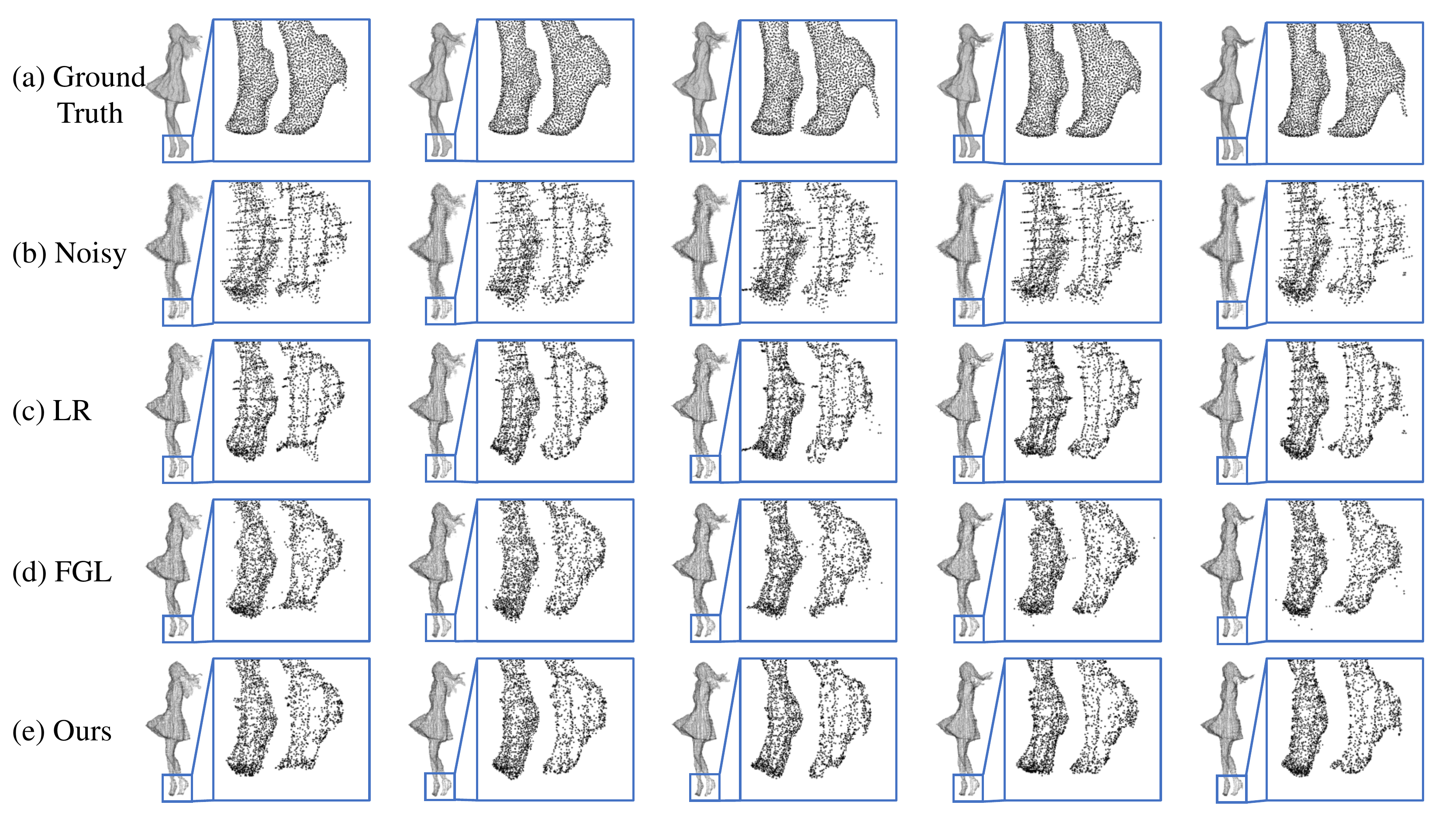}
    \caption{Comparison results from simulated scanner noise ($\sigma=0.01$) for \textit{Redandblack}: (a) The ground truth; (b) The noisy point cloud ($\text{GPSNR}=10.6664$ dB); (c) The denoised result by LR ($\text{GPSNR}=15.2638$ dB); (d) The denoised result by FGL ($\text{GPSNR}=14.2492$ dB); (e) The denoised result by our algorithm ($\text{GPSNR}=16.3530$ dB). Colors are not shown for clear demonstration of geometry denoising.}
    \label{fig:redandblack_scanner}
\end{figure*}

Due to the lack of real-world noisy datasets of dynamic point clouds, we simulate the noise produce by LiDAR sensors and demonstrate the denoising performance in Fig.~\ref{fig:redandblack_scanner}. 
We see that, the denoising results of LR are sometimes distorted in the structure and still noisy along the contour, while the results of FGL are less distorted but suffer from outliers. 
In comparison, our results preserve the structure with cleaner contours, and almost avoid outliers. We also present the quantitative results of the three methods in the caption of the figure, where we achieve the highest GPSNR. 
This demonstrates the potential of our method on real-world noise.





\section{Conclusion}
\label{sec:conclude}
We propose 3D dynamic point cloud denoising by exploiting the temporal consistency between surface patches that correspond to the same underlying manifold, based on a spatial-temporal graph representation. 
Inspired by that graph operators are discrete counterparts of functionals on  Riemannian manifolds, we define a manifold-to-manifold distance and its discrete counterpart on graphs to measure the variation-based intrinsic distance between surface patches in adjacent frames.  
We then construct an initial spatial-temporal graph, where the temporal graph connectivity is based on the manifold-to-manifold distance and the spatial graph connectivity captures the spatial adjacency relations. 
We jointly optimize the desired point cloud and underlying graph representation regularized by both the spatial smoothness and temporal consistency, and reformulate the optimization to design an efficient algorithm. 
Experimental results show that our method significantly outperforms independent denoising of each frame from state-of-the-art static point cloud denoising approaches, on both Gaussian noise and simulated LiDAR noise. 



\bibliographystyle{IEEEtran}

\end{document}